\documentclass[12pt]{article}
\usepackage{epsfig}
\begin{document}

\begin{flushright}
{\tt hep-th/0304180}
\end{flushright}

\vspace{3mm}

\begin{center}
{{{\Large\bf Tachyon Kinks}}\\[12mm]
{Chanju Kim}\\
{\it Department of Physics, Ewha Womans University,
Seoul 120-750, Korea}\\
{\tt cjkim@ewha.ac.kr}\\[5mm]
{Yoonbai Kim~~ and~~ Chong Oh Lee}\\
{\it BK21 Physics Research Division and Institute of 
Basic Science,\\
Sungkyunkwan University, Suwon 440-746, Korea}\\
{\tt yoonbai@skku.ac.kr~~~cohlee@newton.skku.ac.kr}
}
\end{center}

\vspace{5mm}

\begin{abstract}
We study Born-Infeld type tachyonic effective action of unstable
D2-brane with a runaway potential and find rich structure of
static regular solitonic solutions.
There exists only periodic array of tachyon kink-antikinks
in pure tachyonic theory, however, in the presence of electromagnetic fields,
solutions include periodic arrays, topological tachyon kinks, half kink, and
bounces. Computed tension of each kink or single unit of the periodic array
has $T_{1}=\sqrt{2}\pi T_{2}$ or that with a multiplicative factor 
depending on electric field. 
When both electric and magnetic fields are turned on, fundamental string 
charge density has a confined component in addition to a constant piece.
These evidences imply that the obtained codimension-1 objects
are likely to be interpreted as D1-brane (and D1F1) or array of
D1$\bar{{\rm D}}1$ (and D1F1-$\bar{{\rm D}}1$F1) as was the case without
the electromagnetic fields.
Generalization to unstable D$p$-branes is straightforward.
\end{abstract}


\newpage

\setcounter{equation}{0}
\section{Introduction}

Physics of an unstable D-brane or a system of D$\bar{{\rm D}}$ is
manifested through the existence of tachyonic mode. The dynamics of decay
of such unstable branes is described by tachyon 
condensation in the effective theory~\cite{oSen}.
An efficient language to describe the decay and creation of unstable
D$p$-brane
is to study S(pacelike)-brane~\cite{GSt,CGGe,HHW}. Specific
examples are the rolling tachyon~\cite{Sen,MZ} and its family carrying
electromagnetic fields~\cite{MS,RS,KKKK}. 
Since they are time-dependent but spatially homogeneous classical 
configurations in open string theory, they are mainly used for
application to various cosmological issues including inflation, dark matter, 
and reheating~\cite{tacos1}. 
Actual decay process, however, should involve spatial 
inhomogeneity~\cite{CKL,Has} and most of the approaches attempted
dynamical formation of topological kink which is a candidate of D-brane
of codimension one~\cite{CFM}. An obstacle to generate
such topological kink is so-called caustics that kink solution meets 
unavoidable singularity at a finite time irrespective of initial
conditions, e.g., an ordinary kink, a periodic sinusoidal array.

If we want to understand the whole dynamical decay process of an unstable 
D$p$-brane, a relevant question at the moment is to understand viable
form of states after the tachyon is condensed. 
They cannot be perturbative excitations like tachyons or electromagnetic
waves~\cite{Sen,IU}
since all the perturbative degrees of freedom living on the unstable 
D$p$-brane cannot survive 
any more once the D$p$-brane decays away.
In the context of 
open string theory, they should include D$(p-1)$-branes, fundamental 
strings (F1's), and their hybrids with various 
codimensions~\cite{Wit,Yi,Klu,MZ2,AIO,LS,Senk} and small fluctuations on 
them~\cite{MS,KKKK}.

In case of D$(p-1)$-brane, it is described by the topological 
kink~\cite{Klu,MZ2,AIO,LS,Senk} in the effective theory of tachyon 
field~\cite{Gar,Klu,GS,KN,Gar1,Oku}. Recent observation on this kind of
kink~\cite{Senk}
is noteworthy. Static topological kink in the Born-Infeld type tachyon
action with a runaway potential is singular but has finite energy and
tension. The world-volume theory of massless modes on this kink is again
the Born-Infeld type action without any higher derivative corrections.
Inclusion of fermions to this world-volume action leads to restoration of 
supersymmetry and $\kappa$-symmetry so that the obtained kink is identified 
by a BPS D$(p-1)$-brane.

In this paper we will address an indispensable question
related with the aforementioned kink~\cite{Senk}: Can we obtain a static
kink solution without singularity, which reproduces properly its singular 
limit? In the pure tachyonic theory of Born-Infeld type effective action 
with a runaway potential, the unique static regular solution is periodic array
of tachyon kink-anitkinks. In the limit of vanishing pressure, it goes to
that of singular topological tachyon kink-antikinks.
Once Born-Infeld electromagnetism is turned on, the spectrum of regular 
solutions becomes rich. In addition to periodic array, there exist single 
topological tachyon kinks, tachyon half-kink connecting stable and 
unstable vacua, and tachyon bounces specified by the values of 
electromagnetic fields and pressure orthogonal to the soliton direction. 
Without or with electric field less
than critical value orthogonal to the kink direction, tension of a topological
kink or single unit of the periodic array has expected value, 
$T_{1}=\sqrt{2}\pi T_{2}$ for superstring theory. When the electric field
orthogonal to the kink direction is larger than the critical value or its
transverse component is turned on, there is an additional multiplicative 
factor given as a function of the electric field. In the presence of
the longitudinal component of the electric field, F1 charge density has a 
confined component
and its functional form is exactly the same as those of energy density
and transverse component of pressure. Proper singular limit is always 
taken since the energy density of them is given by a $\delta$-function 
or sum of $\delta$-functions in the singular limit 
of those objects. Particularly, for the regular topological tachyon kink with
critical value of electromagnetic fields, all the nice analytic
properties claimed in approximate form only for the singular 
tachyon kink~\cite{Senk} are saturated without any approximation.
The obtained periodic array of 
kink-anitkinks and topological tachyon kink can be interpreted as candidates of
array of D1${\bar{{\rm D}}}$1 (D1F1-${\bar{{\rm D}}}$1F1) and
D1-brane (D1F1), however interpretation of half-kink and tachyon bounces
are not clear, yet. Direct string computation is also lacked at the present 
stage.

We have now many static regular solutions. Although some of
them may presumably be unstable, it is still worthwhile to study
the precise nature of these configurations. 
Another intriguing question is on the small fluctuations on stable objects, 
i.e., the study of the worldvolume action of zero modes and possible
existence of supersymmetry.
Our approach is based on effective field theory so the obtained results
should be understood in terms of string (field) theory~\cite{Senop,ST,RS2}.
Finally it would be quite interesting to investigate the role of our 
solutions in understanding dynamical process of D-brane decays. More 
realistic picture in this direction should also contain closed string 
degrees, e.g, gravitational field and tower of massive closed string 
modes~\cite{OS,CLL,Sen2,LP,LLM}.

The rest of the paper is organized as follows. In section 2, we consider
Born-Infeld type action of a tachyon with $1/\cosh(T/T_{0})$ potential and 
show that a periodic array of kink-antikinks is the unique static regular 
solution. In section 3, we turn on electric field orthogonal to the kink
and find a regular topological kink in addition to a periodic array.
The tension of the kink of codimension one is computed.
In section 4, general form of electromagnetic fields are added to unstable
D2-brane case. The obtained configurations constitute D1F1
in the form of a periodic array, kinks, half-kink, and bounces. Confinement of 
string charge density along the kink is achieved.
We conclude with a summary of the obtained results in section 5.

\setcounter{equation}{0}
\section{Array of Tachyon Kink}
In this section we consider static 
solutions~\cite{Klu,MZ2,AIO,LS,Senk} in pure tachyon 
model described by the following Born-Infeld type effective 
action~\cite{Gar,Klu,Sen,KN}
\begin{eqnarray}\label{act}
S = -T_{p}\int d^{p+1}x\;V(T)\sqrt{-\det (\eta_{\mu\nu}+\partial_{\mu}T
\partial_{\nu}T)}\; ,
\end{eqnarray}
where $T_{p}$ is the tension of the D$p$-brane.
Here we use a runaway tachyon potential
\begin{equation}\label{V2}
V(T)=\frac{1}{\cosh(T/T_{0})},
\end{equation}
which is derived from open string theory~\cite{KN} and was originally
introduced in Ref.~\cite{KKKK,KKKK2,LP,LLM}. In the context of string field theory,
form of the action and the potential is different from our choice~\cite{GS},
which is also allowed due to scheme dependence. $T_{0}$ in the tachyon 
potential has 2 for the bosonic string and $\sqrt{2}$ for the non-BPS D-brane
in the superstring.

To obtain static extended objects of codimension one, we assume that 
the tachyon field depends only on $x=x_{1}$ coordinate such as $T=T(x)$. 
Then, the system is governed by the only nontrivial $x$-component of 
energy-momentum conservation
\begin{equation}\label{tem}
T_{11}'(\equiv\partial_{1}T_{11})=0,
\end{equation}
where nonvanishing components are
\begin{eqnarray}
T_{11}&=&-T_{p}\frac{V(T)}{\sqrt{1+T'^{2}}}<0,\label{t11}\\
T_{ab}&=&-T_{p}V(T)\sqrt{1+T'^{2}}\,\eta_{ab},~~
({\rm diag}(\eta_{ab})=(-1,1,1,...,1),~a,b=0,2,3,...,p).
\label{tab}
\end{eqnarray}

\begin{figure}[t]
\centerline{\psfig{figure=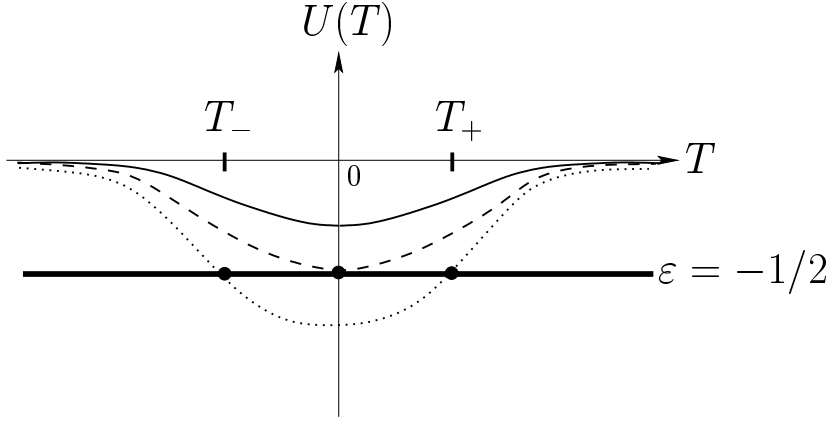,height=60mm}}
\caption{Shapes of 
$U(T)=-\frac{T^2_{p}}{2T^2_{11}}{\rm sech}^2(\frac{T}{T_{0}})$
for various values of $-T_{11}/T_{p}$.}
\label{fig1}
\end{figure}

Since Eq.~(\ref{tem}) forces constant negative pressure $p_{1}=T_{11}$ along 
$x$-direction, we can summarize our system as 
\begin{equation}\label{tpa}
-\frac{1}{2}=\frac{1}{2}T'^{2}-\frac{1}{2}\left
[\frac{T_{p}V(T)}{-T_{11}}\right]^{2}.
\end{equation}
Suppose that we identify Eq.~(\ref{tpa}) as conservation 
of mechanical energy ${\cal E}$ of a hypothetical Newtonian particle 
in one-dimensional motion. Then this hypothetical particle has 
mechanical energy ${\cal E}=-1/2$, unit mass $m=1$, position $T$ at 
time $x$, and is influenced by conservative 
force from potential $U(T)=-(T_{p}V/T_{11})^{2}/2$. 
Therefore possible motions are classified by the value of 
$-T_{11}/T_{p}$
which changes shape (particularly minimum value) of the potential
$U(T)$: (i) When $-T_{11}/T_{p}>1$, no
motion is allowed (see solid curve of $U(T)$ in Fig.~\ref{fig1}). (ii) When 
$-T_{11}/T_{p}=1$, the hypothetical particle stops at $T=0$ eternally (see 
dashed curve of $U(T)$ in Fig.~\ref{fig1}). 
(iii) When $-T_{11}/T_{p}<1$, it oscillates 
between $T_{+}=T_{0}\,{\rm arccosh}(T_{p}/T_{11})$ and 
$T_{-}=-T_{0}\,{\rm arccosh}(T_{p}/T_{11})$ (see dotted curve of $U(T)$ in 
Fig.~\ref{fig1}). (iv) In the limit of 
$-T_{11}/T_{p}\rightarrow 0^{+}$ with keeping all other quantities, 
$T_{\pm}$ approaches positive or negative infinity, respectively.

\begin{figure}[t]
\centerline{\epsfig{figure=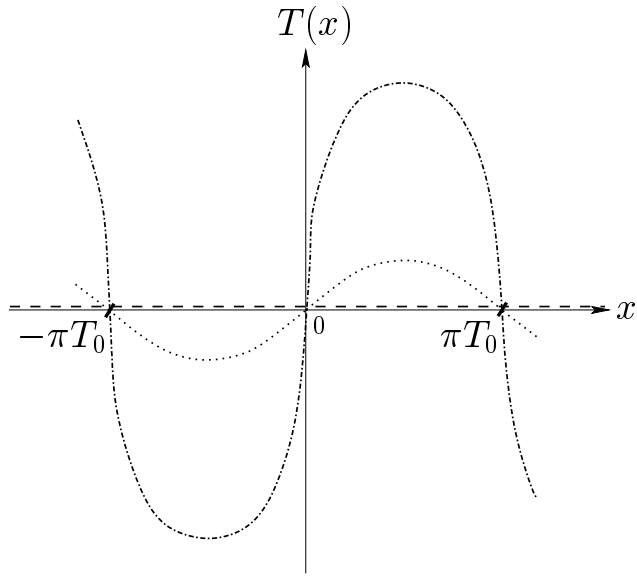,height=80mm}}
\caption{Profiles of tachyon field $T(x)$ for various $-T_{11}/T_{p}$.}
\label{fig2}
\end{figure}

The obtained configurations are interpreted as follows: 

(ii) When the
negative pressure $-T_{11}$ reaches a critical value 
$T_{p}$, tachyon configuration
has a constant value $T=0$ at the unstable vacuum (see dashed line in 
Fig.~\ref{fig2}). 

\begin{figure}[t]
\centerline{\epsfig{figure=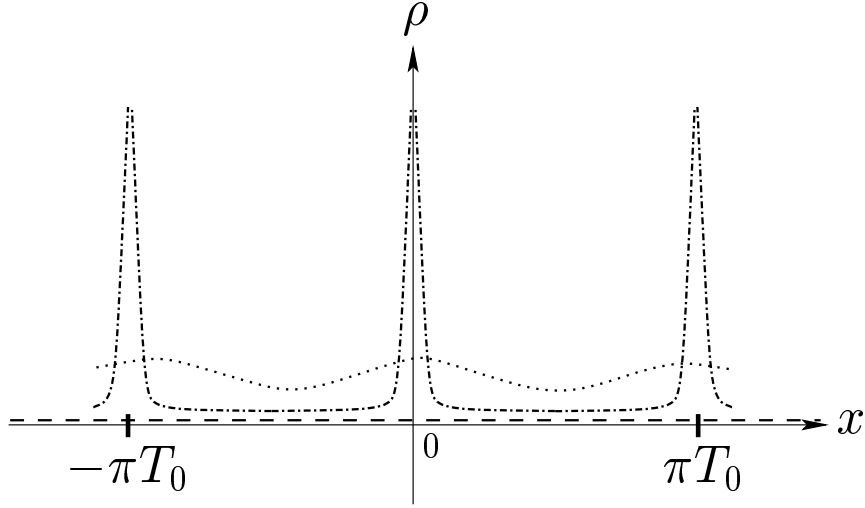,height=70mm}}
\caption{Profiles of energy density $\rho(x)$ and $-T_{22}=-T_{33}=
\cdots=-T_{pp}$.}
\label{fig3}
\end{figure}

(iii) When 
the negative pressure $-T_{11}$ is smaller than the critical value $T_{p}$, 
spatial inhomogeneity is turned on along the $x$-direction in a form 
of a kink, which is expressed by
\begin{eqnarray}
x=\pm\int^{T}_{0}\frac{dT}{\sqrt{(\frac{T_{p}}{T_{11}})^2{\rm sech}^2(T/T_{0})
-1}},
\label{tcf}
\end{eqnarray}
which gives
\begin{equation}\label{tso}
T(x)= T_{0}\;{\rm arcsinh}\left[\sqrt{\left(-\frac{T_{p}}{T_{11}}
\right)^{2}-1}\;\sin\left(\frac{x}{T_{0}}\right)\right].
\end{equation}
The corresponding energy density is given by
\begin{eqnarray}\label{tes}
\rho\equiv T_{00}&=& T_{p}V(T)\sqrt{1+T'^{2}} \nonumber\\
&=&
\frac{-T_{p}^{2}/T_{11}}{1+ \left[(\frac{T_p}{T_{11}})^2-1\right]
\sin^2(x/T_{0})}.
\end{eqnarray}
The solution shows
oscillating behavior between $T_{+}$ and $T_{-}$ with period
$2\pi T_{0}$ which is independent of $T_{11}$
(see dotted curves in Fig.~\ref{fig2} and \ref{fig3}).
Though it may be presumably unstable under a small perturbation,
it is still remarkable that static nonsingular inhomogeneous solution does
exist in the pure tachyon theory.

One may attempt to identify this solution as an
array of kinks and antikinks. Then by integrating the energy density
over a half period of the solution, we find the energy of a kink,
\begin{equation}
T_{p-1}\equiv \int_{-\frac{\pi}{2}T_{0}}^{\frac{\pi}{2}T_{0}}
dx \, T_{00} = \pi T_{0} T_{p},
\label{ttq}
\end{equation}
which can be obtained either by using the expression (\ref{tes}) directly
or by replacing $T'$ in $\rho$ with the help of Eq.~(\ref{tpa}),
\begin{equation}
T_{p-1}= -\frac{T_{p}^{2}}{T_{11}}
\int_{T_{-}}^{T_{+}}dT\frac{V^{2}}{\sqrt{(-T_{p}V/T_{11})^{2}-1}}.
\label{tto} \\
\end{equation}
Note that Eq.~(\ref{ttq}) is independent of the values of $T_\pm$ with the form
of the potential $V(T) = 1/\cosh(T/T_0)$. In fact, it is nothing but the 
tension of the BPS kink identified as
$D(p-1)$-brane~\cite{Senk}. Therefore this solution may be interpreted as
representing an array of $D(p-1)\,\bar D(p-1)$'s.

(iv) In the limit of vanishing pressure $-T_{11}/T_{p}\rightarrow 0^{+}$, 
the period of the static tachyon configuration remains to be the same 
constant $2\pi T_{0}$ but profile of $T(x)$ in Eq.~(\ref{tso}) 
changes abruptly at kink or antikink sites. Accordingly, 
the energy density $\rho(x)$ and all other pressure 
components orthogonal to $x$-direction are more sharply localized. 
They have a peak value at site of the kink, 
$T_{00}(0)=-T_{22}(0)=-T_{33}(0)=\cdots =-T_{pp}(0) =-T_{p}/T_{11}$, and 
decreases to zero exponentially (see dashed, dotted, 
dotted-dashed
lines in Fig.~\ref{fig3} which correspond to the cases (ii), (iii), and (iv),
respectively). Eventually, the solution forms an array of step functions 
\begin{equation}
T(x) \simeq 
T_{0}\;{\rm arcsinh}\left[-\frac{T_{p}}{T_{0}T_{11}}\sin \left(
\frac{x}{T_{0}}\right)\right]
\end{equation}
with an infinite gap
\begin{equation}
T_{+}-T_{-}\simeq \lim_{-T_{11}/T_{p}\rightarrow 0}2T_{0}\ln
\left(-\frac{T_{p}}{T_{0}T_{11}}\right)\rightarrow \infty .
\end{equation}
The energy density (\ref{tes}) then becomes
\begin{equation}
\rho(x)\simeq \pi T_{0}T_{p}\sum_{n=-\infty}^{\infty}\delta(x-n\pi T_{0}).
\end{equation}
In this limit, the kinks become topological~\cite{Klu,MZ2,AIO,LS,Senk},
and develops a singularity of step function with 
infinite gap at each site of kink or antikink.
Note also that the formula of the tension (\ref{tto}) can be approximated as
\begin{equation}\label{tnf}
T_{p-1}\simeq T_{p}\int^{\infty}_{-\infty}dT\, V(T),
\end{equation}
which coincides with that obtained in Ref.~\cite{Senk} for a single 
tachyon kink with singularity.

In this section we showed that the only static regular configuration is 
a periodic array configuration of kink-antikinks. 
In its singular limit, each kink (or antikink) becomes
topological, of which energy density is given by a $\delta$-function, 
however its tension $T_{p-1}$ remains to be a constant.
In the previous approaches~\cite{CFM}, time dependent 
kink configurations have been taken into account mostly by using initial 
configurations like ordinary kink or a periodic sinusoidal array 
for both implication to cosmological 
perturbation or obtaining rolling tachyons with 
inhomogeneity.
Those solutions seem to be suffered by encountering of
unavoidable singularity at a finite time so-called caustics.
Probably it is intriguing to study the instability of aforementioned array
configuration (\ref{tso}) under a small perturbation during time evolution 
where $-T_{11}$ is an adjustable parameter for preparation of an initial
configuration.

\setcounter{equation}{0}
\section{Regular Kink with Electric Field}
In this section, we demonstrate that there exists a static, regular, 
tachyon kink solution when the electric field is larger than or 
equal to the critical value.

As the simplest case, here we will only examine the case of unstable 
D2-brane of which Born-Infeld type action of a tachyon $T$ and an
Abelian gauge field $A_{\mu}$ is
\begin{eqnarray}\label{act1}
S = -T_{2}\int d^3x\;V(T)
\sqrt{-\det (\eta_{\mu\nu}+\partial_{\mu}T\partial_{\nu}T
+F_{\mu\nu})}\; ,
\end{eqnarray}
where $T_{2}$ is tension confined on the D2-brane. The generalization
to higher dimensions should be straightforward.

Let us introduce a few notations ${\bar \eta_{\mu\nu}} = \eta_{\mu\nu} 
+ \partial_{\mu}T\partial_{\nu}T$, ${\bar \eta} \equiv 
\det({\bar{\eta}_{\mu\nu}})$, and ${\bar  F_{\mu\nu}} =  F_{\mu\nu}$.
Define $X_{\mu\nu} \equiv {\bar\eta_{\mu\nu}} + 
{\bar F_{\mu\nu}}$ and $X \equiv \det(X_{\mu\nu})$. 
{}From the definition, $X$ is simplified to
\begin{equation}\label{detX}
X = {\bar \eta}\left(1 + \frac{1}{2}{\bar F_{\mu\nu}}
{\bar F^{\mu\nu}}\right),
\end{equation}
where the transformation rule of a contravariant barred field strength
tensor ${\bar F^{\mu\nu}}$ is 
\begin{equation}\label{trans}
{\bar F^{\mu\nu}} = {\bar \eta^{\mu\alpha}}{\bar \eta^{\nu\beta}}
F_{\alpha\beta},\;\;\;
{\bar \eta}^{\mu\nu} = \eta^{\mu\nu} - \frac{\partial^{\mu}T\partial^{\nu}T}
{1 + \partial_{\rho}T\partial^{\rho}T}.
\end{equation}
Equations of motion derived from the action~(\ref{act1}) are
\begin{eqnarray}\label{eomt}
\partial_{\mu}\left(\frac{V}{\sqrt{-X}}C^{\mu\nu}_{{\rm S}}
\partial_{\nu}T\right) + \sqrt{-X}\frac{dV}{dT} &=& 0 ,\\
\label{eomg}
\partial_{\mu}\left(\frac{V}{\sqrt{-X}}C^{\mu\nu}_{{\rm A}}\right) &=& 0 .
\end{eqnarray}
Here $C^{\mu\nu}_{{\rm S}}$ and $C^{\mu\nu}_{{\rm A}}$ are symmetric and 
antisymmetric parts of the cofactor, respectively
\begin{equation}\label{metc}
C^{\mu\nu} = {\bar \eta}(\bar \eta^{\mu\nu} - 
{\bar F}^{*\mu}{\bar F}^{*\nu}
+ {\bar F^{\mu\nu}})
=C^{\mu\nu}_{{\rm S}} + C^{\mu\nu}_{{\rm A}},
\end{equation}
where 
barred dual field strength tensor of 1-form $\bar{F}^{\ast\mu}$ is 
\begin{equation}
\bar{F}^{\ast\mu}=\frac{\bar{\epsilon}^{\mu\nu\rho}}{2}
\bar{F}_{\nu\rho}~~(\bar{\epsilon}^{\mu\nu\rho}=
\frac{\epsilon^{\mu\nu\rho}}{\sqrt{-\bar{\eta}}},~\epsilon^{012}=-1).
\end{equation}
Energy-momentum tensor is
\begin{eqnarray}\label{emt2}
T_{\mu\nu}
= -T_{2}\frac{V(T)}{\sqrt{-X}}\left[-\eta_{\mu\nu}X + 
\frac{1}{2}\Big(C_{\mu\rho}(\partial_{\nu}T\partial^{\rho}T 
+ F_{\nu}^{\;\rho}) + C_{\nu\rho}(\partial_{\mu}T\partial^{\rho}T 
+ F_{\mu}^{\;\rho})\Big)\right],
\end{eqnarray}
where $C_{\mu\nu} = \eta_{\mu\alpha}\eta_{\nu\beta}C^{\alpha\beta}$.

\begin{figure}[t]
\centerline{\psfig{figure=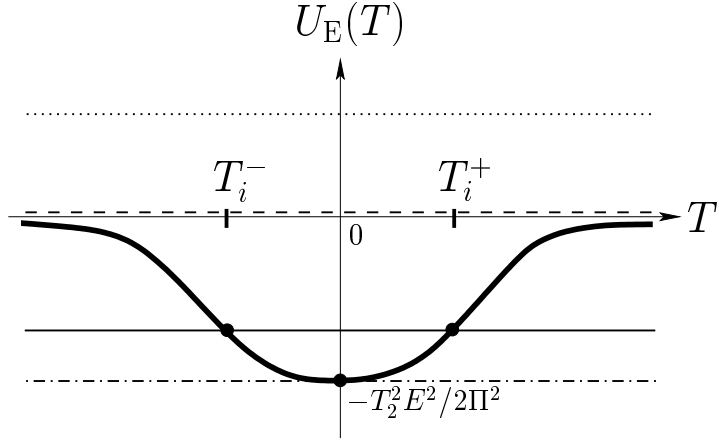,height=60mm}}
\caption{Shape of $U_{{\rm E}}(T)$. 
}
\label{fig4}
\end{figure}

Suppose all fields are static. Since our final goal is to
obtain a straight kink configuration, we assume that the tachyon field
depends only on the $x$-direction  $T=T(x)$.
In this section, for simplicity, we will also consider only the case 
${\bf E}=E({\bf x})\hat{{\bf x}}$ and $B=0$.
General case will be considered in the next section.
Then Bianchi identity $\partial^{\mu}F^{\ast}_{\mu}=0$ which is nothing
but three-dimensional analogue of Faraday's law
\begin{equation}\label{max}
\frac{\partial{B}}{\partial{t}} = -\epsilon_{0ij}\partial_{i}E_{j},
\qquad (E_{i} = F_{i0},~B= \epsilon_{0ij}F_{ij}/2)
\end{equation}
implies $E=E(x)$. The equations for the gauge field (\ref{eomg})
result in constancy of conjugate momentum $\Pi$
\begin{equation}\label{coeq}
\Pi' = 0, 
\end{equation}
where $\Pi$ and $X$ (\ref{detX}) are
\begin{eqnarray}
\frac{\Pi}{E}&=&T_2\frac{V}{\sqrt{-X}}\ge 0,
\label{Pi}\\
-X &=& 1-E^2+T'^{2}\ge 0.
\label{Xx}
\end{eqnarray}
Conservation of the energy-momentum tensor $\partial_{\mu}T^{\mu\nu}=0$
reduces to constant pressure along $x$-direction, $-T_{11}'=(\Pi/E)'=0$, 
so that the electric field $E$ itself 
is a constant. 
It means that the 
solutions are classified by two independent parameters, $(\Pi,E)$
or equivalently $(-T_{11},E)$. 
In the context of string theory, the existence of the constant background
electric field $E$ and its conjugate momentum density $\Pi$ is interpreted
as that of a string fluid consisting of straight F1's along 
$x$-axis~\cite{MS,RS,KKKK}.
Here we assume positive $E$, $\Pi$, and
$-T_{11}$ for convenience without loss of generality.
Therefore, from Eqs.~(\ref{Pi})--(\ref{Xx}), 
we obtain a first-order
differential equation for $T$, consistent with the tachyon equation
(\ref{eomt}),
\begin{equation}\label{ste}
{\cal E}_{{\rm E}}=\frac{1}{2}T'^{2}+U_{{\rm E}}(T),
\end{equation}
where
\begin{eqnarray}
{\cal E}_{{\rm E}}&=&-\frac{1}{2}(1-E^{2}), \label{cale}\\
U_{{\rm E}}(T)&=&-\frac{T_{2}^{2}E^{2}}{2\Pi^{2}}V(T)^{2}
=-\frac{T^{2}_{2}E^{2}}{2\Pi^{2}}\frac{1}{\cosh^{2}(T/T_{0})}.
\label{upo}
\end{eqnarray}
(See Fig.~\ref{fig4} for a schematic shape of $U(T)$.)
Note that for static configurations the energy density (\ref{emt2}) becomes
\begin{eqnarray}\label{een}
\rho\equiv T_{00}=T_{2}\frac{V}{\sqrt{-X}}(1+T'^{2})
=\Pi E+\frac{E}{\Pi}\left[T_{2}V(T)\right]^{2},
\end{eqnarray} 
and pressure $T_{22}$ orthogonal to the configuration is
\begin{equation}\label{epe}
p_{2}\equiv T_{22}=-\frac{\Pi}{E}(1-E^{2}+T'^{2})=-\frac{E}{\Pi}
[T_{2}V(T)]^{2}\le 0.
\end{equation}
Non-constant piece of the energy density (\ref{een}) coincides with
$y$-component of the pressure (\ref{epe}) with opposite signature.
Constant piece of the energy density (\ref{een}) is proportional to
$\Pi$ but the second term given by square of the tachyon potential
is inversely proportional to $\Pi$ so that constant piece dominates in
large $\Pi$ limit and vice versa in small $\Pi$ limit. For the latter,
the pressure orthogonal to the configuration also becomes large.

\begin{figure}[t]
\centerline{\epsfig{figure=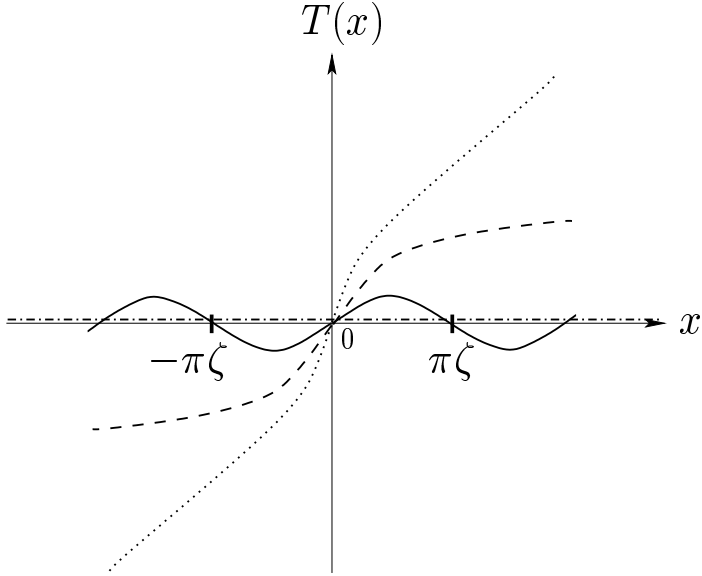,height=80mm}}
\caption{Profiles of tachyon field $T(x)$.
}
\label{fig5}
\end{figure}

\begin{figure}[t]
\centerline{\epsfig{figure=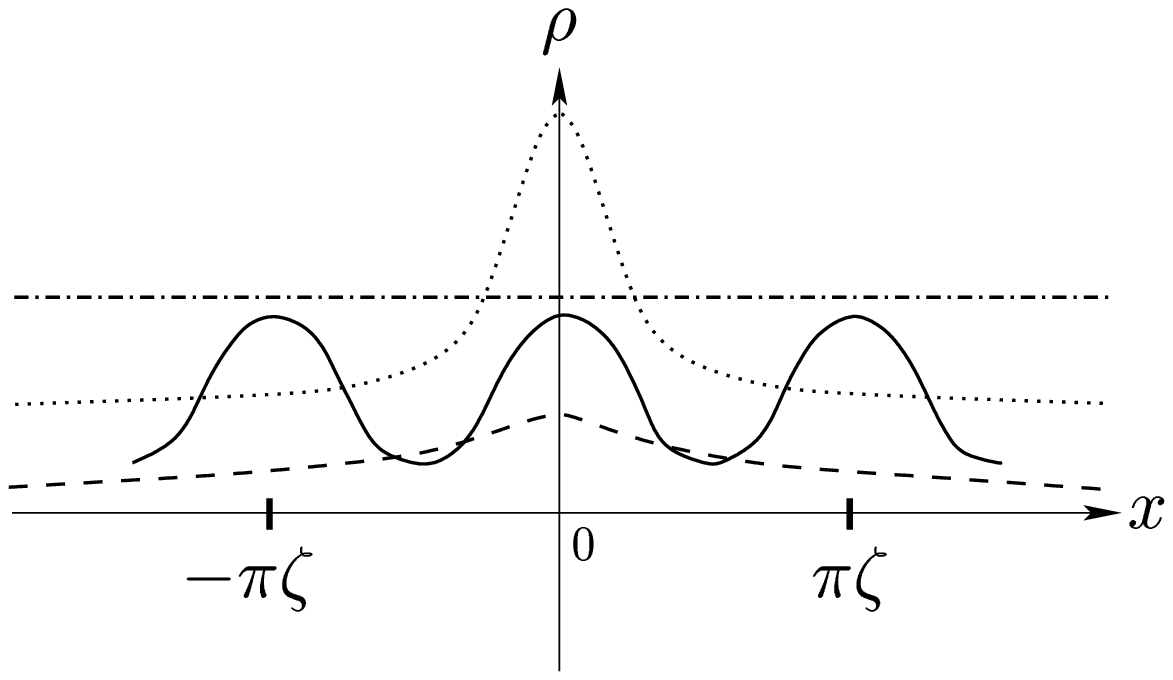,height=70mm}}
\caption{Profiles of energy density $\rho(x)$.
}
\label{fig6}
\end{figure}

Tachyon configurations determined by Eqs.~(\ref{ste})--(\ref{upo}) are
classified by the value of ${\cal E}_{{\rm E}}$. When  
${\cal E}_{{\rm E}}<U(0),$ $(E^{2}>1/[1+(T_{2}/\Pi)^{2}])$, no real 
tachyon configuration is allowed. 
When ${\cal E}=U(0),$ (i.e., $E^{2}=1/[1+(T_{2}/\Pi)^{2}]$; see 
dotted-dashed line in Fig.~\ref{fig4}), 
we have the ontop solution $T(x)=0$ with the corresponding energy 
density  $\rho=\Pi E\left[1+(T_{2}/\Pi)^{2}\right]$ (the dotted-dashed 
lines in Fig.~\ref{fig5} and \ref{fig6}).
For $U_{{\rm E}}(0)<{\cal E}<0,$ ($1/[1+T_{2}^{2}/\Pi^{2}]<E^{2}<1$; 
solid line in 
Fig.~\ref{fig4}), we obtain spatially periodic inhomogeneous configuration
similar to that (\ref{tcf}) in the previous section,
\begin{equation}
x =\int_{0}^{T}\frac{dT}{\sqrt{
E^{2}[1 + T_2^2/\Pi^2 \cosh^{2}(T/T_{0})]-1}},
\end{equation}
which gives
\begin{equation}\label{ear}
T(x) = T_0 \;{\rm arcsinh} \left[
           \sqrt{\frac{E^2T_2^2}{\Pi^2(1-E^2)} - 1} 
	   \;\sin\left(\frac{\sqrt{1-E^2}}{T_0} x \right) \right].
\end{equation}
The tachyon field oscillates between the ``turning points" in Fig.~\ref{fig4},
$T_{i}^{\pm}=\pm T_{0}\,{\rm arccosh}\left(
\frac{T_2 E}{\Pi \sqrt{1-E^{2}}}\right)$ with period 
$2\pi\zeta=2\pi T_0/\sqrt{1-E^2}$. This solution is the generalization of
the array of kinks found in Sec.~2 in the presence of electric field 
in the transverse direction of kinks.
Comparing Eq.~(\ref{ear}) with $E=0$ case (\ref{tso}), we find that
turning on the electric field increases the period and also the effective
tension $T_2 \rightarrow T_2/\sqrt{1-E^2}$.
This kind of phenomenon was expected from the beginning through a rescaling
of $x$-coordinate in the effective action (\ref{act1})
\begin{eqnarray}
S&=&-T_{2}\int dt\, dx\, dy\, V(T)\sqrt{1-E^{2}+\left(\frac{dT}{dx}
\right)^{2}}\nonumber\\
&=&-T_{2}\int dt\, d(\sqrt{1-E^{2}}x)\, dy\, V(T)
\sqrt{1+\left[\frac{dT}{d(\sqrt{1-E^{2}}x)}\right]^{2}}.
\end{eqnarray}
Substituting the solution (\ref{ear}) into Eq.~(\ref{een}), we have 
\begin{eqnarray}\label{err}
\rho -\Pi E = -p_{2} = \frac{E}{\Pi}
\frac{T_{2}^{2}}{1+\left[\frac{E^{2}T_{2}^{2}}{\Pi^{2}(1-E^{2})}-1\right]
\sin^{2}\left(x/\zeta\right)}.
\end{eqnarray}
The tachyon profile $T=T(x)$ and the energy density $\rho=\rho(x)$
for this periodic solution are represented as the solid lines in
Fig.~\ref{fig5} and \ref{fig6}.

As in Sec.~2 we integrate the energy density of the kink, i.e. the
localized piece of $\rho(x)$, over the half period to get its tension 
($p=2$ in the
present case)
\begin{eqnarray}
T_{p-1}&=&\frac{ET_{2}^{2}}{\Pi}\int_{-\frac{\pi}{2}\zeta}^{\frac{\pi}{2}\zeta}
dx \, V^{2}(T(x)) \nonumber\\
&=&\pi T_{0}T_{p}.\label{etp}
\end{eqnarray}
It is exactly the same value as the tension $T_{p-1}$ 
without electric field (\ref{ttq}) and
again independent of the ``turning points'' $T_{i}^{\pm}$. Moreover it
is independent of the electric field $E$ or F1 charge density $\Pi$
in the transverse direction.
In the limit of $\Pi\rightarrow 0$ with fixed $E$, which corresponds to
$-T_{11}\rightarrow 0$, both the energy density (\ref{een}) and the pressure
(\ref{epe}) are given by sums of $\delta$-functions
\begin{eqnarray}
\rho(x)\simeq  T_{p-1}\sum_{n=-\infty}^{\infty}\delta
(x-n\pi T_{0}/\sqrt{1-E^{2}})
\simeq -p_{2}.
\end{eqnarray}

As ${\cal E}$ approaches $0^{-}$, i.e., when $E^{2}$ approaches 1 (the 
dashed line in Fig.~\ref{fig4}), $T_{i}^{\pm}$ stretches to infinity
and we obtain new types of solutions, which does not exist in the limit of
vanishing electric field. In fact the solution becomes
a regular static single-kink configuration with $T'(\pm\infty)=0$ 
(the dashed line in Fig.~\ref{fig5}),
\begin{equation}\label{exs}
T(x)=T_{0}\;{\rm arcsinh}\left(\frac{T_{2}}{\Pi T_{0}}x\right).
\end{equation}
This regularity can also be understood through localization of the action 
(\ref{act1}) as follows
\begin{eqnarray}
S&=&-T_{p}\int dt\, dx\,d^{p-1}y\, V(T)\sqrt{1-E^{2}+T'^{2}}\nonumber\\
&\stackrel{E=\pm 1}{=}&-(\pm)T_{p}\int dt\, dx\, d^{p-1}y\, V(T) T'
\label{exy}\\
&=&-\int dt\,d^{p-1}y\;T_{p}\int_{-\infty}^{\infty}dT\, V(T),
\label{ext}
\end{eqnarray}
where $+$ ($-$) in the second line (\ref{exy}) 
corresponds to the kink (the antikink). The exact integral formula for
the tachyon field in the third line
(\ref{ext}) is nothing but that of the tension $T_{p-1}$ (\ref{tnf})
obtained through rather complicated manipulation only for the singular 
kink~\cite{Senk}.
In this case, the energy density $\rho$ is given in a particularly simple 
form,
\begin{equation}\label{ed0}
\rho-\Pi = -p_{2} = \pi T_0 T_2 \cdot
\frac{\xi/\pi}{x^2 + \xi^2},
\end{equation}
where $\xi \equiv \Pi T_0/T_2$ represents the width of the kink.
Note that as $\Pi$ goes to zero the localized energy density approaches
a $\delta$-function with energy $T_{p-1}=\pi T_0 T_p$, 
while large $\Pi$ broadens
the width.

{}Finally, when ${\cal E}_{{\rm E}}>0$, ($E^{2}>1$; see the dotted line in 
Fig.~\ref{fig4}), the solution is given by 
\begin{equation}\label{eki}
T(x) = T_0 \;{\rm arcsinh} \left[
           \sqrt{1+\frac{E^{2}T_{2}^{2}}{\Pi^{2}(E^2-1)}}
	   \sinh\left(\frac{\sqrt{E^2-1}}{T_0} x \right) \right],
\end{equation} 
which has a finite asymptotic slope $T'(\pm\infty)\neq0$.
The energy density (\ref{een}) for this solution also has a constant
and a localized piece which coincides with the pressure $p_{2}$ with
opposite signature
\begin{equation}
\rho(x)-\Pi E =-p_{2} = \frac{ET_{2}^{2}}{\Pi}
\frac{1}{1+\left[1+\frac{E^{2}T_{2}^{2}}{\Pi^{2}(E^{2}-1)}\right]
{\rm sinh}^{2}\left(\frac{\sqrt{E^{2}-1}}{T_{0}}x\right)}.
\end{equation}
Similar to the previous case, we obtain
\begin{eqnarray}
T_{p-1}&=&\frac{ET_{p}^{2}}{\Pi}\int_{-\infty}^{\infty}dx V^{2}(T(x))
\nonumber\\
&=& 2T_{0}T_{p}\;{\rm arctan}\left(\frac{ET_{2}}{\Pi\sqrt{E^{2}-1}}
\right)
\label{etk}
\end{eqnarray}
which is no longer a constant and is less than $\pi T_{0}T_{p}$.
Of course, $E\rightarrow 1^{+}$ limit with fixed $\Pi$ reproduces trivially
the previous case of topological kink (\ref{exs}) and thereby its tension
$T_{p-1}$ approaches $\pi T_{0}T_{p}$. 
The profiles of $T(x)$ and the energy density $\rho(x)$ are plotted
as dotted lines in Fig.~\ref{fig5} and Fig.~\ref{fig6}.
As ${\cal E}\rightarrow 0^{+}$, the slope
at spatial infinity $T'(\pm\infty)$ goes to zero as expected. For huge
electric field limit $E\rightarrow \infty$, the tachyon kink becomes
sharply peaked, $T'(\pm\infty)\rightarrow \pm \infty$,
with keeping finite tension. 

All the array and kinks are specified by two independent constants,
$x$-component of electric field $E$ and conjugate momentum $\Pi$ for a given
D2-brane.
In synthesis for fixed $\Pi$, as the electric field $E$ increases,
irregularity is induced in both energy density and transverse component of 
the pressure and becomes sharply peaked, and its pattern changes from
an array to single kink.

\setcounter{equation}{0}
\section{Confined F1 Charge along the Tachyon Kink}
In this section we consider the most general configuration of the
static electromagnetic fields with $x$-dependence alone
\begin{equation}\label{2em}
{\bf E}=E_{1}(x)\hat{{\bf x}}+E_{2}(x)\hat{{\bf y}},\qquad
B=B(x).
\end{equation}
Since the system of our interest is still unstable D2-brane, we can use
the formulas in Eqs.~(\ref{act1})--(\ref{emt2}). If we insert Eq.~(\ref{2em})
into the Faraday's law (\ref{max}), it forces constancy of $y$-component of the 
electric field $E_{2}$. Conjugate momenta $\Pi_{i}$ 
of the gauge fields are
\begin{eqnarray}
\Pi_{1}&=&T_{2}\frac{V(T)}{\sqrt{-X}}E_{1},
\label{2e1}\\
\Pi_{2}&=&T_{2}\frac{V(T)}{\sqrt{-X}}(1+T'^{2})E_{2}.
\label{2e2}
\end{eqnarray}
{}From time-component of the gauge equations (\ref{eomg}), we see that
$\Pi_{1}$ should be a constant, while $\Pi_{2}$ need not be. Note 
that $\Pi_2$ is nonzero only when the parallel component of the
electric field $E_2$ is turned on.
Now the $y$-component of the energy-momentum conservation 
$\partial_{1}T^{12}=0$ is automatically satisfied and the 
$x$-component, $\partial_{1}T^{11}=0$, gives
\begin{equation}\label{2tv}
\frac{\Pi_{1}}{E_{1}}=T_{2}\frac{V(T)}{\sqrt{-X}}={\rm a~positive~constant},
\end{equation}
so that $x$-component of the electric field $E_{1}$ is also a constant.
The $y$-component of the gauge equations (\ref{eomg}) dictates
constancy of the magnetic field $B$. Therefore, 
all the electromagnetic fields $({\bf E},B)$ are actually
independent of $x$.
Then the remaining time-component of the energy-momentum conservation 
is also satisfied automatically: 
$\partial_{1}[ (T_{2}V/\sqrt{-X})(E_{2}+E_{1}B)B]=0$.

Eliminating non-constant $\Pi_{2}$ in the remaining two equations
(\ref{2e1})--(\ref{2e2}), we can again summarize dynamics of our system 
by a single first-order equation as done in the previous sections
\begin{equation}\label{2st}
{\cal E}_{{\rm EM}}=\frac{1}{2}T'^{2}+U_{{\rm EM}}(T),
\end{equation}
where
\begin{eqnarray}
{\cal E}_{{\rm EM}}
&=&-\frac{1-{\bf E}^{2}+B^{2}}{2(1-E_{2}^{2})}, \label{2cal}\\
U_{{\rm EM}}(T)&=&-\frac{T_{2}^{2}E_{1}^{2}}{2\Pi_{1}^{2}(1-E_{2}^{2})}V(T)^{2}
=-\frac{T^{2}_{2}E_{1}^{2}}{2\Pi_{1}^{2}(1-E_{2}^{2})}\frac{1}{
\cosh^{2}(T/T_{0})}.
\label{2up}
\end{eqnarray}
In the limit of $E_{2}\rightarrow 0$ and $B\rightarrow 0$, it is consistent
with Eqs.~(\ref{ste})--(\ref{upo}). Therefore, all the solutions are
classified by a set of four parameters, i.e., $(\Pi_{1},E_{1},E_{2},B)$
or $(T_{11},E_{1},E_{2},B)$ where 
\begin{equation}
-T_{11}=\frac{\Pi_{1}}{E_{1}}(1-E_{2}^{2}),
\end{equation}
which has sign flip at the critical value of $E_{2}=\pm 1$.

In this setup, some components of the energy-momentum tensor are nonvanishing 
constants,
\begin{eqnarray}
T_{0i}&=&-\frac{\Pi_{1}}{E_{1}}\epsilon_{0ij}E_{j}B, \nonumber \\
T_{12}&=&-\frac{\Pi_{1}}{E_{1}}E_{1}E_{2}.
\end{eqnarray}
The other components of the the energy-momentum tensor have nontrivial 
$x$-dependence:
\begin{eqnarray}
\rho\equiv T_{00} &=&T_{2}\frac{V}{\sqrt{-X}}(1+T'^{2}+B^{2}) \nonumber \\
&=&\frac{\Pi_{1}(E_{1}^{2}-B^{2}E_{2}^{2})}{E_{1}(1-E_{2}^{2})}
+\frac{E_{1}}{\Pi_{1}(1-E_{2}^{2})}[T_{2}V(T)]^{2} \nonumber\\
&\equiv&\rho_{\rm c} + \rho_{\rm l},
\label{2ee} \\
p_{2}\equiv T_{22}&=&-T_{2}\frac{V}{\sqrt{-X}}(1+T'^{2}-E_1^{2}) \nonumber\\
&=&-\frac{\Pi_{1}(E_{1}^{2}E_{2}^{2}-B^{2})}{E_{1}
(1-E_{2}^{2})}-\frac{E_{1}}{\Pi_{1}(1-E_{2}^{2})}[T_{2}V(T)]^{2} \nonumber\\
&\equiv&p_{\rm 2c} + p_{\rm 2l}.
\label{2p2}
\end{eqnarray} 
Note also that the string charge density $\Pi_2$ in the $y$-direction has 
essentially the same $x$-dependence,
\begin{eqnarray}\label{2con}
\Pi_{2}&=&\frac{\Pi_{1}E_{2}(E_{1}^{2}-B^{2})}{E_{1}(1-E_{2}^{2})}
+\frac{E_{1}E_{2}}{\Pi_{1}(1-E_{2}^{2})}[T_{2}V(T)]^{2} \nonumber \\
&\equiv & \Pi_{\rm 2c} + \Pi_{\rm 2l}.
\end{eqnarray}
These three inhomogeneous quantities, namely energy density (\ref{2ee}), 
pressure (\ref{2p2}), and string charge density (\ref{2con}) in $y$-direction
share a few properties which may be related to confinement of D1F1: 
(i) They are composed of a constant and a common $x$-dependent part,
$\Pi_{\rm 2l}(x) = E_2 \rho_{\rm l}(x) = -E_2 p_{\rm 2l}(x)$.
(ii) The constant term is proportional to $\Pi_{1}$ but the localized piece
is inversely proportional to $\Pi_{1}$. (iii) They have an overall
multiplicative factor $1/(1-E_{2}^{2})$
so that they flip signature at the critical value $E_{2}=\pm 1$.

Now let us study solution spectra of Eqs.~(\ref{2st})--(\ref{2up}) 
in what follows. First, note that the coefficient of the potential term in
Eq.~(\ref{2up}) changes the sign at $E_2^2=1$. Therefore we will separately
examine the system according to the value of $E_2$ (we assume $E_2\ge 0$
with no loss of generality): $0\le E_2<1$, $E_2=1$ and $E_2>1$.

When $E_{2}$ is smaller than one, there is not much to do since 
Eq.~(\ref{2st}) is essentially identical to that in the previous section.
With some trivial replacements of parameters, we obtain the following result.
(See Figs.~\ref{fig5} and \ref{fig6} for the possible types of solutions.)

(i) When $E_{2}<1$ and $1-{\bf E}^{2}+B^{2}>T_{2}^{2}E_{1}^{2}/\Pi_{1}^{2}$,
no solution exists. 

(ii) When $E_{2}<1$ and 
$1-{\bf E}^{2}+B^{2}=T_{2}^{2}E_{1}^{2}/\Pi_{1}^{2}$,
there is only a constant solution $T(x)=0$.

(iii) When $E_{2}<1$ and 
$0<1-{\bf E}^{2}+B^{2}<T_{2}^{2}E_{1}^{2}/\Pi_{1}^{2}$, we have the solution
of kink-antikink array,
\begin{equation}
T(x)=T_{0}\;{\rm arcsinh}\left[
\sqrt{u^{2}-1}
\;\sin\left(\frac{x}{\zeta_B}
\right)\right],
\end{equation}
where $u$ and $\zeta_B$ are defined by
\begin{eqnarray}
u^{2}&=&\frac{E_{1}^{2}T_{2}^{2}}{\Pi_{1}^{2}|1-{\bf E}^{2}+B^{2}|},
\label{2uu}\\
\zeta_B&=&\frac{T_0}{\sqrt{2{|\cal E}_{\rm EM}|}}
= \sqrt{\left|\frac{1-E_{2}^{2}}{1-{\bf E}^{2}+B^{2}}\right|}\; T_{0}.
\label{2zt}
\end{eqnarray}
Then the square of the tachyon potential $V(T)^{2}$, to which the localized
pieces of $\rho$, $p_2$ and $\Pi_2$ are proportional, is given by
\begin{equation}
V^2(T(x)) = \frac{1}{1+(u^{2}-1)\sin^{2}(x/\zeta_B)}.
\end{equation} 
{}From integration of the localized part of the energy density $\rho(x)$ 
along $x$-axis, we obtain the tension $T_{1}$ of codimension one object 
\begin{equation}
T_{1}=\frac{\pi T_{0}T_{2}}{\sqrt{1-E_{2}^{2}}},
\end{equation}
which is larger than the previous case (\ref{etp}) by a multiplicative 
factor $1/\sqrt{1-E_{2}^{2}}$. This is expected since the energy density
of a Born-Infeld theory increases precisely by this factor when a constant
electric field is turned on on the world volume. 

In the limit of $E_{2}\rightarrow 1$, $T_{1}$ diverges and oscillation
becomes rapid $\zeta_B\rightarrow 0$ with infinite peak of the energy density. 
Let us take another limit $\Pi_{1}\rightarrow 0$ which leads to
$u\rightarrow \infty$ but $\zeta_B$ remains finite.
Then $V(T)^{2}$ is written by a sum of $\delta$-functions so that we have
\begin{eqnarray}
\rho(x) &\simeq& \frac{\pi T_{0}T_{2}}{\sqrt{1-E_{2}^{2}}} 
\sum_{n=-\infty}^{\infty}\delta(x-2\pi\zeta_B),
\label{2r5}\\ 
\Pi_{2}&\simeq& \frac{\pi T_{0}T_{2}E_{2}}{\sqrt{1-E_{2}^{2}}} 
\sum_{n=-\infty}^{\infty}\delta(x-2\pi\zeta_B).
\label{2p5}
\end{eqnarray}
We can read the tension of each kink (or antikink) 
$\pi T_{0}T_{2}/\sqrt{1-E_{2}^{2}}$ 
from Eq.~(\ref{2r5}) and string charge density of each kink 
$\pi T_{0}T_{2}E_{2}/\sqrt{1-E_{2}^{2}}$ from Eq.~(\ref{2p5}).

(iv) When $1-{\bf E}^{2}+B^{2}\rightarrow 0$, both $u$ and $\zeta_B$ 
diverge with finite ratio $\zeta_B/u=\Pi_{1}/T_{2}E_{1}$. 
Since $V(T)^{2}$ takes a Lorentzian shape
\begin{equation}
V^{2}(T)\rightarrow \pi\frac{\zeta_B}{u}
\frac{\zeta_B/\pi u}{x^{2}+(\zeta_B/u)^{2}},
\end{equation}
so do the localized pieces of energy density (\ref{2ee}), transverse pressure
(\ref{2p2}), and string charge density (\ref{2con}). This case corresponds
to a single-kink solution as in the previous section.
The configuration of tachyon kink is given by
\begin{equation}
T(x)=T_{0}\;{\rm arcsinh}\left(\frac{x}{\zeta_B}\right).
\end{equation}
In this limit, the action (\ref{act1}) is rewritten again in a localized form 
\begin{eqnarray}
S&=&-T_{p}\int dt\, dx\,dy\, V(T)\sqrt{1-{\bf E}^{2}+B^{2}+(1-E_{2}^{2})T'^{2}}
\nonumber\\
&\stackrel{1-{\bf E}^{2}+B^{2}=0}{=}&-(\pm)\sqrt{1-E_{2}^{2}}\;
T_{p}\int dt\, dx\, dy\, V(T) T' \label{exy2}\\
&=&-\int dt\,dy
\sqrt{1-E_{2}^{2}}\; T_{p}\int_{-\infty}^{\infty}dT\, V(T),
\label{ext2}
\end{eqnarray}
where $+$ ($-$) in the second line (\ref{exy2}) 
corresponds to the kink (the antikink). In the third line (\ref{ext2}) we
obtain the same formula of tension (\ref{tnf}) with the aforementioned
overall factor
$\sqrt{1-E_{2}^{2}}$.

(v) Finally, when $E_2<1$ and $1-{\bf E}^2 = B^2 <0$, we obtain the solution
similar to Eq.~(\ref{eki}). The details are omitted.

Now we consider the case $E_{2}=1$. In this case the solution is trivial
since $T'$ disappears, for example, in Eq.~(\ref{2st}) or in the tachyon 
equation of motion (\ref{eomt}). Then the only remaining equation is
$V'(T)=0$. Therefore there are only homogeneous solutions $T=0$ or 
$T=\pm\infty$.

\begin{figure}[t]
\centerline{\psfig{figure=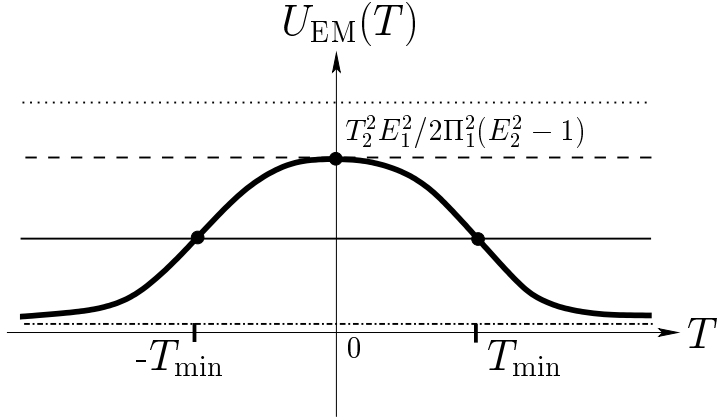,height=60mm}}
\caption{Shape of $U_{{\rm EM}}(T)$ when $|E_{2}|>1$. 
}
\label{fig7}
\end{figure}

As mentioned before, once the magnitude of $E_{2}$
is larger than 1, the property of our system changes drastically 
because of the sign flip of the potential $U_{{\rm EM}}(T)$ as shown in 
Fig.~\ref{fig7}.
Similar to the analysis of the previous solutions, 
tachyon configurations are classified by 
the value of ${\cal E}_{{\rm EM}}$ (\ref{2cal}).
In fact, the solution configurations should be analogous to
rolling tachyon solutions in unstable D2-brane with electromagnetic 
fields when $1-{\bf E}^{2}+B^{2}\ge 0$.
{}From Eq.~(\ref{act1}) the action of our system is rewritten as
\begin{eqnarray}
S&=&-T_{2}\int dt\, dx\, dy\, V(T)\sqrt{1-{\bf E}^{2}+B^{2}-(E_{2}^{2}-1)
\left(\frac{dT}{dx}\right)^{2}}\nonumber\\
&=&-T_{2}\sqrt{E_2^2-1}\int dt\, 
d\left(\frac{T_{0}x}{\zeta_{B}}\right)\, dy V(T)
\sqrt{1 -\left[\frac{dT}{d(T_{0}x/\zeta_{B})}\right]^{2}}.
\label{222}
\end{eqnarray}
Note that the signature of $x$-direction flips when $E_2 >1$. This action
is actually exactly the same as that of rolling tachyon which is
given by
\begin{eqnarray}
S&=&-T_{2}\int dt\, dx\, dy\, V(T)\sqrt{1-{\bf E}^{2}+B^{2}-(1+B^{2})
\left(\frac{dT}{dt}\right)^{2}}\nonumber\\
&=&-T_{2}\sqrt{1+B^2}\int d\left(\frac{T_{0}t}{\xi}\right)\, dx\, dy 
V(T)\sqrt{1-\left[\frac{dT}{d(T_{0}t/\xi)}\right]^{2}},
\label{2bb}
\end{eqnarray}
where
\begin{equation}
\xi=\sqrt{\frac{1+B^{2}}{1-{\bf E}^{2}+B^{2}}}T_{0}.
\end{equation}
Thus, there exists a one-to-one correspondence between a kink solution
with spatial distribution and the time evolution of a homogeneous 
rolling tachyon solution. With this identification,
the pressure $-T_{11}$, in our system corresponds to 
the Hamiltonian density ${\cal H}$, in the rolling tachyon system.

Now we describe the solutions in detail when $E_2>1$.

\begin{figure}[t]
\centerline{\epsfig{figure=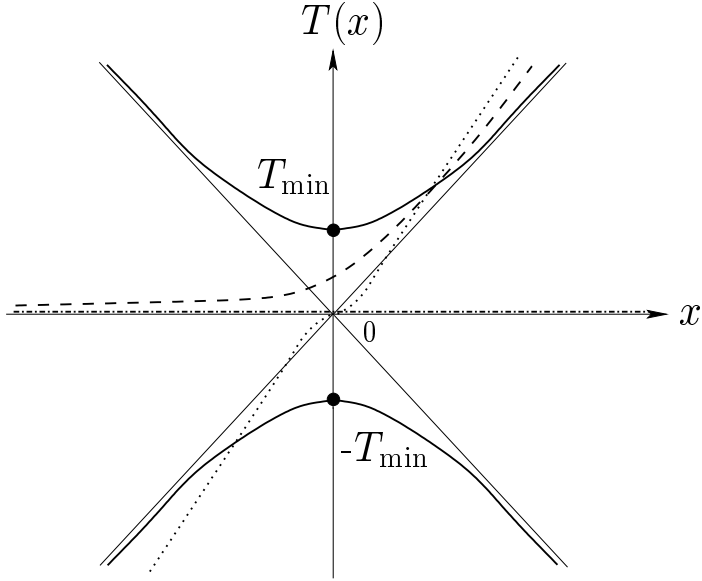,height=80mm}}
\caption{Profiles of tachyon kink and bounce $T(x)$.}
\label{fig8}
\end{figure}

\begin{figure}[t]
\centerline{\epsfig{figure=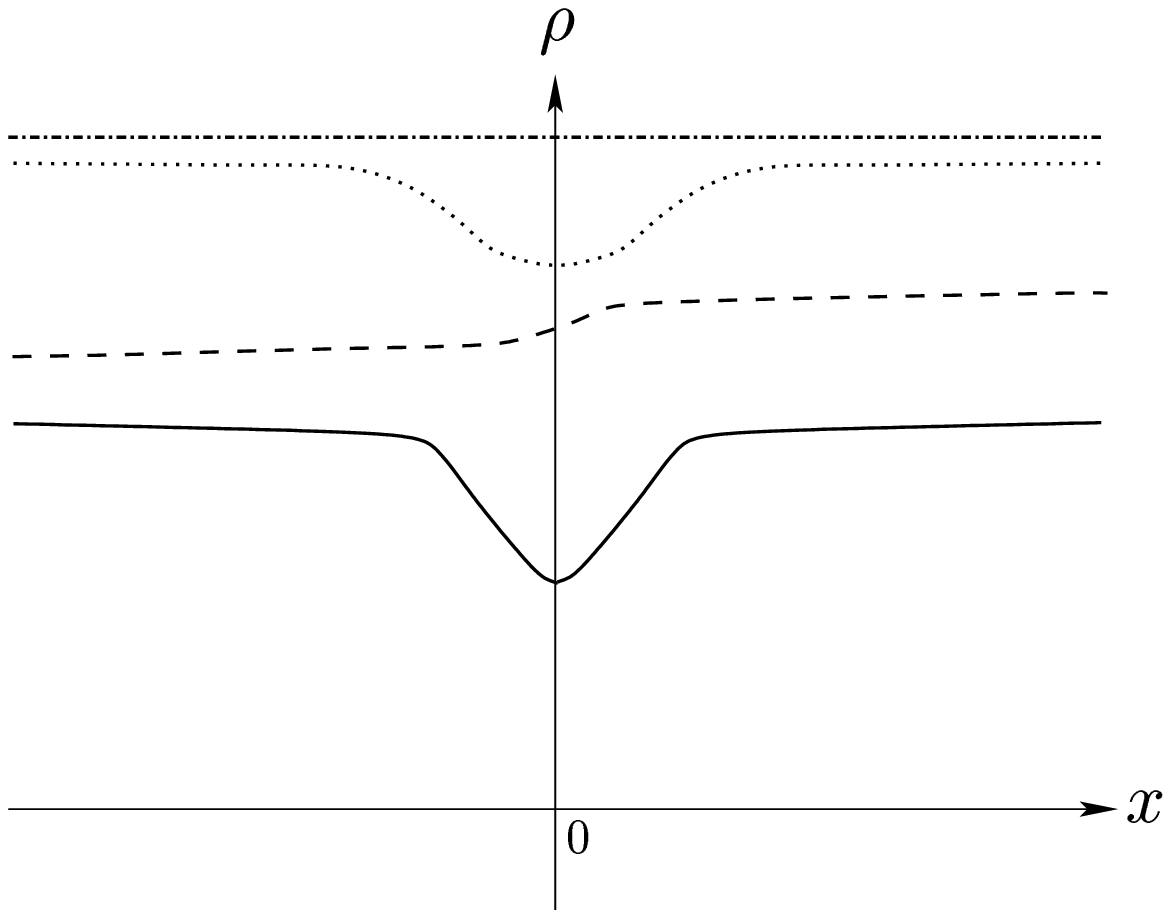,height=70mm}}
\caption{Profiles of energy density $\rho(x)$.}
\label{fig9}
\end{figure}

(i) As ${\cal E}_{{\rm EM}}\rightarrow 0^{+}$,
(i.e, ${\bf E}^{2}-B^{2}\rightarrow1$;
see the dot-dashed lines in Figs.~\ref{fig7}--\ref{fig9}),
static constant vacuum solution is allowed at $T(x)=\pm \infty$.
Since the tachyon potential vanishes for $T=\pm\infty$, the localized terms
in $\rho(x)$, $p_2(x)$, and $\Pi_2(x)$ are all zero.

(ii)
When the energy ${\cal E}_{{\rm EM}}$ is larger than zero but smaller than the
top of tachyon potential
(i.e., $1-(T_{2}E_{1}/\Pi_{1})^{2}<{\bf E}^{2}-B^{2}<1$; see the solid line
in Fig.~\ref{fig7}), there is a turning point $T_{\rm min}$ such that
\begin{equation}\label{2mi}
|T|(x)\ge T_{\rm min}=T_{0}\, {\rm arccosh}(u).
\end{equation}
As shown by the solid curves in the Fig.~\ref{fig8}, configuration is convex
up
(or convex down) so we will call this solution a tachyon bounce.
Explicit form of the solution is given by
\begin{equation}\label{2xe}
T(x)=T_{0}\;{\rm arcsinh}\left[\sqrt{u^{2}-1}\cosh\left(\frac{x}{\zeta_B}
\right)\right],
\end{equation}
where $u$ and $\zeta_B$ are given in Eqs.~(\ref{2uu}) and (\ref{2zt}),
respectively.
Its asymptotic slopes are
\begin{equation}
T'(\pm\infty)=\pm T_0/\zeta_B,
\end{equation}
which are shown by the two solid straight lines in Fig.~\ref{fig8}.
Note that, since $E_2>1$, the localized parts of $\rho(x)$, $p_2(x)$, and
$\Pi_{2}(x)$ all flip their signs.
For example,
the energy density has positive constant term and a negative localized
contribution near the origin (see the solid line in Fig.~\ref{fig9})
\begin{equation}\label{2na}
\rho =\frac{\Pi_{1}(B^{2}E_{2}^{2}-E_{1}^{2})}{E_{1}(E_{2}^{2}-1)}
-\frac{E_{1}T_2^2}{\Pi_{1}(E_{2}^{2}-1)}
\frac{1}{1+(u^{2}-1)\cosh^{2}(x/\zeta_B)}.
\end{equation} 
Also it means that the transverse component of pressure $p_{2}$ is positive
and the string charge density $\Pi_{2}$ is negative.

(iii) When the ``energy'' ${\cal E}_{{\rm EM}}$ is the same as
the maximum of the ``potential'' $U_{{\rm EM}}(0)$,
(${\bf E}^{2}-B^{2}=1-(T_{2}E_{1}/\Pi_{1})^{2}$; see the dashed line in
Fig.~\ref{fig7}), we obtain the trivial vacuum ontop solution $T(x)=0$.
In addition, there is also the tachyon ``half-kink'' (or ``half-antikink'')
solution which connects the unstable symmetric vacuum $T(-\infty)=0$ and
a stable broken vacuum $T(\infty)=\pm\infty$
(see the dashed curve in Fig.~\ref{fig8}),
\begin{equation}
T(x)=\pm T_{0}\;{\rm arcsinh}\left[\exp \left(\frac{x}{\zeta_B}\right)\right].
\end{equation}
Since the half-kink connects two vacua with different vacuum energy as
$V(T=0)>V(T=\pm\infty)$, the energy density is monotonically increasing
(see the dashed curve in Fig.~\ref{fig9}),
\begin{equation}
\rho(x)=\frac{\Pi_{1}(B^{2}E_{2}^{2}-E_{1}^{2})}{E_{1}(E_{2}^{2}-1)}
-\frac{E_{1}T_{2}^{2}}{\Pi_{1}(E_{2}^{2}-1)}
\frac{1}{1+\exp(2x/\zeta_B)}.
\end{equation}
The transverse component of pressure $p_{2}$ and the string charge density
$\Pi_{2}$ are also monotonic.

(iv) If ${\cal E}_{{\rm EM}}> U_{EM}(0)$,
(${\bf E}^{2}-B^{2}
<1-(T_{2}E_{1}/\Pi_{1})^{2}$, see the dotted line in Fig.~\ref{fig7}),
we have
\begin{equation}\label{2gr}
T(x)=T_0 {\rm arcsinh}\left[ \sqrt{1-u^2} \sinh\left( 
         \frac{x}{\zeta_B}\right) \right].
\end{equation}
Configuration is monotonically increasing (or decreasing) from
$T(-\infty)=\mp\infty$ to $T(\infty)=\pm\infty$ (see also the dotted curve
in Fig.~\ref{fig8}).
Opposite to the similar kink solutions in the previous section, slope of
the solutions has minimum value at the origin,
and maximum at infinity.
This solution can be considered as two half-kink solutions joined at the
origin. The energy density of this solution is given by 
\begin{equation}\label{2rh}
\rho =\frac{\Pi_{1}(B^{2}E_{2}^{2}-E_{1}^{2})}{E_{1}(E_{2}^{2}-1)}
-\frac{E_{1}T_2^2}{\Pi_{1}(E_{2}^{2}-1)}
\frac{1}{1+(1-u^{2})\sinh^{2}(x/\zeta_B)}.
\end{equation}
It is plotted in Fig.~\ref{fig9} with the dotted line.

\setcounter{equation}{0}
\section{Conclusion}
In this paper, static solutions have been investigated in Born-Infeld 
type tachyonic effective action with and without electromagnetic fields.

In pure tachyonic theory, we have shown that the periodic array of tachyon 
kink-antikinks
is the unique static regular solution. In the limit of vanishing pressure 
along the array, the solution becomes an array of step functions 
with an infinite gap. The values of tension of single unit kink (or antikink)
$T_{p-1}=\sqrt{2}\pi T_{p}$ and period remain constant irrespective of the 
value of pressure.

When the electrostatic field orthogonal to the tachyon soliton is turned on 
in an unstable D2-brane, a periodic array,
and regular topological tachyon kinks are obtained, classified by the value
of electric field with fixed conjugate momentum. For a given electric field,
taking limit of vanishing conjugate momentum leads to singular limit where
both energy density and transverse component of pressure are given by sum of 
$\delta$-functions or a $\delta$-function.
When the electric field is smaller than or equal to critical
value, tension is $T_{p-1}=\sqrt{2}\pi T_{p}$. When the electric field is 
larger than the critical value, a multiplicative factor depending on 
the value of electric field appears.

For the general case with both electric and magnetic fields,
spectra of solutions are specified by four parameters, three components of
electromagnetic fields and one pressure component along the soliton,
and divided into two classes by transverse component of
electric field. When it is smaller than critical value, the spectra of 
solitons are exactly the same as the cases of pure electric field, i.e., they 
are periodic array of tachyon kink-antikinks and topological tachyon kinks.
Minor difference appears in the constant scales and the tension.
Major difference is that they carry confined component of F1 charge density
so that each kink is presumably a D1F1.
When the transverse component of electric field is larger than critical 
value, we additionally found half-kink
connecting unstable and stable vacua, tachyon bounce, and topological
tachyon kink.   

The topological tachyon kink at the critical value of the electric field
orthogonal to the tachyon soliton seems to be a BPS object, but it
should be addressed in future works by obtaining its
worldvolume action and checking existing supersymmetry as has been done in
Ref.~\cite{Senk}.

\section*{Acknowledgements}
We would like to thank O-Kab Kwon, Ashoke Sen, and Piljin Yi for valuable
discussions and comments.
This work was supported by Korea Research Foundation Grant
KRF-2002-070-C00025(C.K.) and is
the result of research activities (Astrophysical Research
Center for the Structure and Evolution of the Cosmos (ARCSEC) and
the Basic Research Program, R01-2000-000-00021-0)
supported by Korea Science $\&$ Engineering Foundation(Y.K.).

\end{document}